\documentclass[aps,prd,preprint,superscriptaddress,amsmath,amssymb,showpacs]{revtex4-1}
\usepackage{dcolumn}
\usepackage{graphicx}
\usepackage{float}
\usepackage{physics}
\usepackage[colorlinks=true,allcolors=blue]{hyperref}

\begin{document}

\title{Spin polarization and anomalous magnetic moment in a (2+1)-flavor
Nambu-Jona-Lasinio model in a thermomagnetic background}

\author{Yi-Wei Qiu}
\affiliation{College of Science, China Three Gorges University, Yichang 443002, China}

\author{Sheng-Qin Feng}
\email{Corresponding author: fengsq@ctgu.edu.cn}
\affiliation{College of Science, China Three Gorges University, Yichang 443002, China}
\affiliation{Key Laboratory of Quark and Lepton Physics (MOE) and Institute of Particle Physics,\\
Central China Normal University, Wuhan 430079, China}
\affiliation{Center for Astronomy and Space Sciences, China Three Gorges University, Yichang 443002, China}

\date{\today}

\begin{abstract}
Abstract:  We investigate the magnetized QCD matter and chiral phase transition in a (2+1)-flavor Nambu-Jona-Lasinio (NJL) model at finite temperature and chemical potential by comparing the contributions from the tensor spin polarization (TSP) and anomalous magnetic moment (AMM) of quarks. For light $u$ and $d$ quarks, when TSP and AMM are not considered, the magnetized system is characterized by magnetic catalysis. The introduction of TSP will further enhance the magnetic catalytic characteristics. On the other hand, when AMM is introduced, the phase-transition temperature decreases with the magnetic field, which is the feature of inverse magnetic catalysis. The phase diagram of $u$ and $d$ quarks will change from the crossover phase transition to the first-order phase transition with the increase of magnetic field and chemical potential when AMM is induced. The phase diagram will not change from the crossover phase transition to the first order phase transition when TSP is induced. For the phase diagram of strange $s$ quark, whether TSP or AMM is induced, the phase diagram will keep a crossover phase transition with the increase of magnetic field and chemical potential.
\end{abstract}


\maketitle

\section{Introduction}\label{sec:01_intro}
Comprehending properties of QCD matter under a strong magnetic field is of essential importance to further investigate the evolution of the early Universe \cite{Vachaspati:1991nm}, noncentral heavy-ion collisions \cite{Skokov:2009qp,Deng:2012pc,Mo:2013qya,Zhong:2014cda}, neutron-star mergers \cite{Kiuchi:2015sga,Baiotti:2016qnr}, and the interior of magnetars \cite{Tatsumi:2006wr,Duncan:1992hi}. The exploration of the QCD vacuum and strongly-interacting matter under external strong magnetic fields has attracted much attention, (see reviews, e.g., Refs.~\cite{Bzdak:2019pkr,Kharzeev:2015znc,Huang:2015oca,Andersen:2014xxa,Miransky:2015ava}). Here we stress the study of the magnetic field of non-central heavy-ion collisions, which comes from the laboratory simulations. The magnetic field reaches up to $\sqrt{eB} \sim 0.1\mathrm{GeV}$ for \textrm{RHIC} and $\sqrt{eB} \sim 0.5~\mathrm{GeV}$ for LHC in noncentral heavy-ion collisions. This magnetic field is external since it is generated by the spectators, and though it has a very short lifetime(of the order of 1 $\textrm{fm}/c $). However, as shown in Refs.~\cite{Gursoy:2014aka,She:2017icp,Chen:2019qoe,Chen:2017lsf}, the presence of the quark-gluon plasma (QGP) medium response effect, substantially delays the decay of these time-dependent magnetic fields. This is why in most cases, the effect of constant and
uniform magnetic fields on quark matter is discussed in the literature. The magnetic field coincides with the production of the \textrm{QGP} and thus may have a fairly important effect on the properties of the phase transition such as the chiral magnetic effect (\textrm{CME}) \cite{Kharzeev:2007jp,Fukushima:2008xe,She:2017icp,Guo:2019joy,Deng:2021kyd}, magnetic catalysis (MC) in the vacuum \cite{Gusynin:1999pq,Gusynin:1995nb,Klevansky:1989vi}, inverse magnetic catalysis (\textrm{IMC}) around the chiral phase transition~\cite{Bali:2013esa,Bali:2012zg,Bali:2011qj,DElia:2018xwo}.

The magnetic field can lead to spin polarization, that is, the condensation of quark- antiquark $\left ( \bar{q} q \right ) $ pairs with spin parallel. Reference.\cite{Ferrer:2013noa} shows that a tensor-type interaction $\sim \left ( \bar\psi  \Sigma^{3}  \psi \right)^{2}  +\left ( \bar\psi i\gamma ^{5}  \Sigma^{3}  \psi \right)^{2}$ produces a spin polarization (SP) $\left \langle \bar{\psi} i\gamma ^{1} \gamma ^{2}  \psi  \right \rangle $, which is very similar to the anomalous magnetic moment (AMM) produced by quarks in a magnetic field. The tensor-polarization operator $\bar{\psi}\sigma ^{\mu \nu }  \psi $ can also be named as the spin-polarization operator, or the spin density since $\bar{\psi} \sigma ^{12 }  \psi =\psi\gamma ^{0} \Sigma ^{3} \psi$. If the quark spinor $\psi$ is projected into the subspin space $\psi ={{\psi }_{\uparrow }}+{{\psi }_{\downarrow }}$ corresponding to $\bar{\psi }{{\sigma }^{12}}\psi \sim \left\langle {{{\bar{\psi }}}_{\uparrow }}{{\psi }_{\uparrow }} \right\rangle -\left\langle {{{\bar{\psi }}}_{\downarrow }}{{\psi }_{\downarrow }} \right\rangle $, then this can be used to measure the difference between the spin-up quark pair and the spin-down quark pair.

We investigate the magnetized QCD matter in a (2+1)-flavor Nambu-Jona-Lasinio (NJL) model at finite temperature and chemical potential by comparing the contributions from the tensor spin polarization (TSP) and AMM of quarks. For a particle with charge $e$, mass $m$ and spin $\vec{s}$, its corresponding magnetic moment (MM) is $\mu$ corresponding to the $\bar{q} q $ pair with antiparallel spin pairs, it has a net magnetic moment (MM), so the chiral condensation triggers a dynamic \textrm{AMM}. Under the action of the magnetic field, the net MM tends to be parallel to the magnetic field. For a SP with $\bar{q} q $ pair-parallel spin pairing, the MM of spin-aligned quarks and antiquarks cancel each other, and the spin-polarization pairing does not present a net MM. Therefore, compared with the chiral condensation with a nonzero net MM, the total MM of the system considering SP condensation will reduce. Therefore, systems with a spin polarization are expected to exhibit relative diamagnetism. At high temperatures, the pair of $\bar{q} q $ dissociates, and all charged quarks become a single small magnet, which are arranged in turn along the magnetic field. Therefore, \textrm{QCD} matter at high temperatures manifests paramagnetism.

The catalysis of chiral symmetry breaking induced by a magnetic field, namely the MC effect, can be easily understood from dimension reduction. On the other hand, \textrm{IMC} effect, the critical temperature of the chiral phase transition decreases with the increasing magnetic field, which is intuitively contradictory to the MC effect and is still a puzzle. There are many publications trying to explain \textrm{IMC} by considering running coupling constant generated by the magnetic field \cite{Ferrer:2014qka} and chiral imbalance caused by sphaleron transition or instanton anti-instanton pairing \cite{Chao:2013qpa}. Some interesting and novel properties of magnetized QCD materials have recently been presented by lattice calculations, for example, magnetized materials exhibit paramagnetism (positive susceptibility) at high temperatures and diamagnetism (negative susceptibility) at low temperatures \cite{Bali:2012jv,Bali:2020bcn}.

Recently, the effect of the \textrm{AMM} of quarks has drawn quite a lot of interest  \cite{Fayazbakhsh:2014mca,Ferrer:2015wca,Chaudhuri:2019lbw,Ghosh:2020xwp,Chaudhuri:2020lga,Mao:2018jdo,Mei:2020jzn} in order to investigate the IMC effect. The dynamical chiral symmetry broken is known as one of the most important characteristics of QCD, which makes quarks achieve a dynamical mass of \textrm{QCD}. References \cite{Ferrer:2009nq,Chang:2010hb} pointed out that quarks' AMM can also be dynamically produced like the dynamical quark mass. Therefore, once quarks achieve dynamical mass, they should also achieve dynamical AMM \cite{Ferrer:2008dy,Preis:2010cq,Ferrer:2009nq,Bicudo:1998qb}. The coefficient $\kappa $ of quarks' AMM  in the magnetic field by the effective interaction $\frac{1}{2} q\kappa F_{\mu \nu } \bar{\psi} \sigma ^{\mu \nu }\psi$ ($ \sigma ^{\mu \nu }=\frac{i}{2}\left [ \gamma ^{\mu } ,\gamma ^{\nu }  \right ] $) is introduced, and the \textrm{IMC} effect at finite temperature is proposed by Ref. \cite{Xu:2020yag}.  For \textrm{QCD}, both explicit and spontaneous chiral symmetry breaking are dedicated to the \textrm{AMM} of quarks, which is also called dynamical \textrm{AMM} \cite{Chang:2010hb}.

In this paper, we investigate the magnetism of \textrm{QCD} matter and chiral phase transition with the contributions from the \textrm{TSP} and the \textrm{AMM} of quarks, respectively. This paper is organized as follows: We introduce the (2+1)-flavor \textrm{NJL} models by including the \textrm{AMM} and the \textrm{TSP} in Sec. II.  In order to investigate the \textrm{MC} and \textrm{IMC} features by the \textrm{AMM} and \textrm{TSP}, the dependencies of dynamical mass, entropy, sound-velocity, and critical point on the magnetic field and temperature are studied in Sec. III. Finally, we make the summaries and conclusions in Sec. IV.

\section{THE 2 + 1 FLAVORS NJL MODEL UNDER A MAGNETIC FIELD}\label{sec:02 setup}
The Lagrangian density of the (2 + 1)-flavor \textrm{NJL} model \cite{Buballa:2003qv,Hatsuda:1994pi} in the presence of an external magnetic field is given as
\begin{equation}\label{eq:01}
\begin{split}
 \mathcal{L}=& \bar{\psi }\left( i{{\gamma }^{\mu }}{{D}_{\mu }}+{{\gamma }^{0}}\mu -m \right)\psi + {{G}_{s}}\sum\limits_{a=0}^{8}{\left[ {{\left( \bar{\psi }{{\lambda }_{a}}\psi  \right)}^{2}}+{{\left( \bar{\psi }i{{\gamma }^{5}}{{\lambda }_{a}}\psi  \right)}^{2}} \right]} \\
 & -K \left[ \det \bar{\psi }\left( 1+{{\gamma }_{5}} \right)\psi +\det \bar{\psi }\left( 1-{{\gamma }_{5}} \right)\psi  \right],\\
 \end{split}
\end{equation}
where the quark field $\psi $ carries three flavors ($f=u,\,d,\,s$) and three colors ($c=r,g,b$ ), and ${{\lambda }_{a}}(a=1,\cdots N_{f}^{2}-1)$ represents the SU(3) Gell-Mann matrices in the three flavor space. Current quark mass $m$ is considered as ${{m}_{u}}={{m}_{d}}$ for isospin symmetry of light quarks, strange quark mass ${{m}_{s}}$ is different from the other light quark ($m_u$ and $m_d$) masses. The difference between the strange and nonstrange quark masses obviously breaks the SU(3) flavor symmetry. We assume that the quark chemical potentials of the strange and nonstrange quarks are the same, and take $\mu$ as the quark chemical potential. A covariant derivative with magnetic field is introduced as ${{D}_{u}}={{\partial }_{\mu }}+\operatorname{i}QA_{\mu }^{\operatorname{ext}}$, and the charge matrix in flavor space is

\begin{equation}\label{eq:02}
Q=\operatorname{diag}\left( {{q}_{u}},{{q}_{d}},{{q}_{\text{s}}} \right)=\operatorname{diag}\left( \frac{2}{3},-\frac{1}{3},-\frac{1}{3} \right).
\end{equation}

In general, if one chooses the gauge field $A_{\mu }^{ext}=\left( 0,0,B{{x}_{1}},0 \right)$, the constant magnetic field should point at the ${{x}^{3}}$-direction. The $K$ term of Eq. (1) is the Kobayashi-Maskawa-t'Hooft interaction term \cite{Hatsuda:1994pi,Vogl:1991qt,Rehberg:1995kh}.

\subsection{The introduction of a (2 + 1)- flavors NJL model with TSP}
It is shown that \cite{Ferrer:2013noa,Fayazbakhsh:2014mca} the breaking of the rotational symmetry by a uniform magnetic field induces a separation between longitudinal and transverse-fermion modes along the direction of the magnetic field. This separation gives rise to the effective splitting of the couplings in the one-gluon exchange interactions on which the \textrm{NJL} models are usually based. This splitting is therefore reported in the four-fermion couplings of a \textrm{QCD}-inspired \textrm{NJL} model in a magnetic field, and we can use the Fierz identities in a magnetic field \cite{Ferrer:2013noa,Ferrer:2014qka,Lin:2022ied} to propose the interactions of scalar and tensor of the (2 + 1)-flavor \textrm{NJL} Lagrangian,
\begin{equation}\label{eq:03}
\begin{split}
   {{\mathcal{L}}_{\text{TSP}}}= & \bar{\psi }\left( i{{\gamma }^{\mu }}{{D}_{\mu }}+{{\gamma }^{0}}\mu -m \right)\psi +{{G}_{s}}\sum\limits_{a=0}^{8}{\left[ {{\left( \bar{\psi }{{\lambda }_{a}}\psi  \right)}^{2}}+{{\left( \bar{\psi }i{{\gamma }^{5}}{{\lambda }_{a}}\psi  \right)}^{2}} \right]} + {{G}_{t}}\sum\limits_{a=0}^{8} \\
  & {\left\{ {{\left( \bar{\psi }{{\Sigma }_{3}}{{\lambda }_{a}}\psi  \right)}^{2}}+{{\left( \bar{\psi }{{\Sigma }_{3}}i{{\gamma }^{5}}{{\lambda }_{a}}\psi  \right)}^{2}} \right\}}-K \left\{ \det [\bar{\psi }\left( 1+{{\gamma }_{5}} \right)\psi] +\det [\bar{\psi }\left( 1-{{\gamma }_{5}} \right)\psi]  \right\}.
 \end{split}
\end{equation}

The coupling constant ${{G}_{s}}$ in the scalar/pseudoscalar channel is closely related to the spontaneously chiral symmetry breaking, which produces a dynamical quark mass, and the tensor/ pseudotensor channels term ${{G}_{t}}\sum\limits_{a=0}^{8}{\left[ {{\left( \bar{\psi }_{f}^{c}{{\Sigma }^{3}}{{\lambda }_{a}}\psi _{f}^{c} \right)}^{2}}+{{\left( \bar{\psi }_{f}^{c}i{{\Sigma }^{3}}{{\gamma }^{5}}{{\lambda }_{a}}\psi _{f}^{c} \right)}^{2}} \right]}$ is closely related to the spin-spin interaction, which causes spin- polarization condensation.

For the (2 + 1)-flavor NJL model, tensor-type interaction at the mean field level leads to the two types of spin polarization as
\begin{equation}\label{eq:04}
\begin{split}
& {{F}_{3}}=-2{{G}_{t}}\left\langle \bar{\psi }{{\Sigma }^{3}}{{\lambda }_{3}}\psi  \right\rangle , \\
& {{F}_{8}}=-2{{G}_{t}}\left\langle \bar{\psi }{{\Sigma }^{3}}{{\lambda }_{8}}\psi  \right\rangle.
\end{split}
\end{equation}

In general, $F_{3}$ contains only $u$ and $d$ quark spin polarization condensates, on the other hand, $F_{8}$ is associated with the strange quark spin polarization condensate. The running coupling constants are divided into longitudinal (${{g}_{\parallel }}$) and transverse (${{g}_{\bot }}$) components due to the existence of the magnetic field. In our current study, the couplings of the above \textrm{NJL} interactions relevant to quark-gluon vertex coupling are expressed as ${{G}_{s}}=\left( g_{||}^{2}+g_{\bot }^{2} \right)/{{\Lambda }^{2}}$ and ${{G}_{t}}=\left( g_{||}^{2}-g_{\bot }^{2} \right)/{{\Lambda }^{2}}$.  The distinguishing transverse and parallel Fierz identities automatically create a new channel of a four-fermion interaction term with second-order tensor structure in Lagrangian density during the transformation from splitting quark-gluon coupling to the scalar and pseudoscalar bilinear quantity \cite{Ferrer:2013noa}. ${{G}_{s}}$ and ${{G}_{t}}$ can be considered as the scalar and tensor channel interaction couplings, respectively.

The effective potential of using a standardized process is given
\begin{equation}\label{eq:05}
\begin{split}
 {{\Omega }_{\text{TSP}}}=&{{G}_{s}}\sum\limits_{f=u,d,s}{\left\langle \overline{\psi }\psi \right\rangle }_{f}^{2}+{{G}_{t}}{{\left\langle \overline{\psi }{{\lambda }_{3}}{{\Sigma }^{3}}\psi  \right\rangle }^{2}}+{{G}_{t}}{{\left\langle \overline{\psi }{{\lambda }_{8}}{{\Sigma }^{3}}\psi  \right\rangle }^{2}}
  -\frac{{N}_{c}}{2\pi } \sum\limits_{f=u,d,s}{\left| {{q}_{f}}B \right|}\sum\limits_{l=0}^{\infty }{{{\alpha }_{l}}} \mathop{\int }_{-\infty }^{\infty }\frac{d{{p}_{z}}}{2\pi} \\
  & \times \left\{ {{\varepsilon }_{f,l,\eta }}+ T \ln \left[ 1+\exp \left( \frac{-{{\varepsilon }_{f,l,\eta }}-\mu }{T} \right) \right]
  + T \ln \left[ 1+\exp \left( \frac{-{{\varepsilon }_{f,l,\eta }}+\mu }{T} \right) \right] \right\} \\
 &+ 4K{{\left\langle \overline{\psi }\psi  \right\rangle }_{u}}{{\left\langle \overline{\psi }\psi  \right\rangle }_{d}}{{\left\langle \overline{\psi }\psi  \right\rangle }_{s}}
 \end{split}
\end{equation}
where $l$= 0, 1, 2 ... represents the quantum number of Landau level and $\eta =\pm 1$ corresponds to the two kinds of the spin direction of quark-antiquark ($\bar{q}q$) pair. The contribution of nondegenerate particles due to spin difference at nonlowest Landau energy levels can be taken into account with the definition of this new operator ${{\alpha }_{l}}={{\delta }_{0,l}}+\Delta \left( l \right)\sum\limits_{\eta =\pm 1}{{}}$, where $\Delta \left( l \right)$ is denoted by
\begin{equation}\label{eq:06}
\begin{split}
\Delta \left( l \right)=\left\{ \begin{matrix}
   0  \\
   1  \\
\end{matrix} \right.\quad \quad \begin{matrix}
   l=0  \\
   l>0,  \\
\end{matrix}
 \end{split}
\end{equation}
and the energy spectrum of the lowest Landau level ( LLL) $\left( l=0 \right)$ and non-LLL ($l\ne 0$) are given as
\begin{equation} \label{eq:07}
\begin{split}
 & \varepsilon _{_{u,l=\text{0}}}^{2}=p_{z}^{2}+{{\left( {{M}_{f}}+\left( {{F}_{3}}+\frac{{{F}_{8}}}{\sqrt{\text{3}}} \right) \right)}^{2}}, \\
 & \varepsilon _{_{u,l\ne 0,\eta = \pm 1}}^{2} = p_{z}^{2}+{{\left( \sqrt{{{M}_{f}}^{2}+2|{{q}_{f}}B|l}+\eta \left( {{F}_{3}}+\frac{{{F}_{8}}}{\sqrt{\text{3}}} \right) \right)}^{2}}, \\
 & \varepsilon _{_{d,l=0}}^{2}=p_{z}^{2}+{{\left( {{M}_{f}}+\left( {{F}_{3}}-\frac{{{F}_{8}}}{\sqrt{\text{3}}} \right) \right)}^{2}}, \\
 & \varepsilon _{_{d,l\ne 0,\eta = \pm 1}}^{2}=p_{z}^{2}+{{\left( \sqrt{{{M}_{f}}^{2}+2|{{q}_{f}}B|l}+\eta \left( {{F}_{3}}-\frac{{{F}_{8}}}{\sqrt{\text{3}}} \right) \right)}^{2}}, \\
 & \varepsilon _{_{s,l=0}}^{2}=p_{z}^{2}+{{\left( {{M}_{f}}+\left( \frac{2{{F}_{8}}}{\sqrt{\text{3}}} \right) \right)}^{2}}, \\
 & \varepsilon _{_{s,l\ne 0,\eta=\pm 1 }}^{2}=p_{z}^{2}+{{\left( \sqrt{{{M}_{f}}^{2}+2|{{q}_{f}}B|l}+\eta \left( \frac{2{{F}_{8}}}{\sqrt{\text{3}}} \right) \right)}^{2}}.\\
\end{split}
\end{equation}

Note that the breaking of energy-spectrum degeneracy caused by spin is known as the Zeeman effect. Therefore, the contributions of spin come not only from the ground state of the Landau levels, but also from the whole excited states of the Landau levels.
The tensor-condensate parameter ${{F}_{3}}$ and ${{F}_{\text{8}}}$ are self-consistently satisfied the minimum of the thermodynamic potential, which are similar to dynamical quark mass ${{M}_{f}}$. At first, one can obtain three gap equations for ${{M}_{f}}$ ($f = u, d, s$)
\begin{equation}\label{eq:08}
\frac{\partial {{\Omega }_{\text{TSP}}}\left( {{M}_{f}},{{F}_{3}},{{F}_{8}} \right)}{\partial {{M}_{f}}}=0,
\end{equation}

\noindent  and the other two gap equations for $F_{3}$ and $F_{8}$ are given as
\begin{equation}\label{eq:09}
\begin{split}
& \frac{\partial {{\Omega }_{\text{TSP}}}\left( {{M}_{f}},{{F}_{3}},{{F}_{8}} \right)}{\partial {{F}_{3}}} = 0, \\
& \frac{\partial {{\Omega }_{\text{TSP}}}\left( {{M}_{f}},{{F}_{3}},{{F}_{8}} \right)}{\partial {{F}_{8}}} = 0 . \\
\end{split}
\end{equation}

To ensure that the thermodynamic potential in vacuum returns to zero, we define the normalized thermodynamic potential as the effective potential
\begin{equation}\label{eq:010}
{{\Omega }_{\text{eff}}}\left( T,\mu ,eB \right)=\Omega \left( T,\mu ,eB \right)-\Omega \left( 0,0,eB \right).
\end{equation}


Some of the relevant thermodynamical quantities can be evaluated by the effective potential. The quark number density is

\begin{equation}\label{eq:011}
{{\rho }_{f}}=\sum\limits_{l,\eta}{\frac{{{N}_{c}}\left| {{q}_{f}}eB \right|}{\text{4}{{\pi }^{2}}}}\int\limits_{\text{-}\infty }^{\infty }{d{{p}_{z}}}\left( {{n}^{+}}-{{n}^{-}} \right)\ ,
\end{equation}
where ${{n}^{\pm }}={1}/{\left( \exp \left[ \left( {{\varepsilon }_{f, l,\eta}}\mp \mu  \right)/T \right]+1 \right)} $ is quark (antiquark) number distribution. The entropy density $S_{f} = -\frac{\partial {{\Omega }_{\text{eff}}}}{\partial T}$ is given as
\begin{equation}\label{eq:012}
\begin{split}
  S_{f} = -\sum\limits_{l,\eta}{\frac{{{N}_{c}}\left| {{q}_{f}}eB \right|}{\text{4}{{\pi }^{2}}}}\int\limits_{\text{-}\infty }^{\infty }{d{{p}_{z}}} \left[ \ln \left( 1-{{n}^{+}} \right)+ \ln \left( 1-{{n}^{-}} \right)-\frac{{\varepsilon_{f,l,\eta}}}{T}\left( {{n}^{+}}+{{n}^{-}} \right)\text{+}\frac{\mu }{T}({{n}^{+}}-{{n}^{-}}) \right].
\end{split}
\end{equation}

The energy density is given as
\begin{equation}\label{eq:013}
\varepsilon =T\frac{\partial P}{\partial T}\text{+}\mu \frac{\partial P}{\partial \mu }-P,
\end{equation}
where $P$ is pressure. The square of sound-speed is defined as
\begin{equation}\label{eq:014}
c_{s}^{2}=\frac{\partial P}{\partial \varepsilon }={{\left( \frac{\mu }{{{S}_{f}}}\frac{\partial {{\rho }_{f}}}{\partial T}\text{+}\frac{T}{{{S}_{f}}}\frac{\partial {{S}_{f}}}{\partial T} \right)}^{\text{-1}}}.
\end{equation}

\subsection{the introduction of the (2 + 1)- flavor NJL model with AMM}
The effective Lagrangian density of the (2 + 1)- flavor with \textrm{AMM} \cite{Buballa:2003qv,Hatsuda:1994pi} is given as
\begin{equation}\label{eq:015}
	\begin{split}
	& {{\mathcal{L}}_{\text{AMM}}}=\bar{\psi }\left( i{{\gamma }^{\mu }}{{D}_{\mu }}+{{\gamma }^{0}}\mu -m\text{+}\frac{\text{1}}{\text{2}}{{q}_{f}}\kappa {{\sigma }^{\mu \nu }}{{F}_{\mu \nu }} \right)\psi  \\
	& +{{G}_{s}}\sum\limits_{a=0}^{8}{\left[ {{\left( \bar{\psi }{{\lambda }_{a}}\psi  \right)}^{2}}+{{\left( \bar{\psi }i{{\gamma }^{5}}{{\lambda }_{a}}\psi  \right)}^{2}} \right]}-K\left[ \det \bar{\psi }\left( 1+{{\gamma }_{5}} \right)\psi +\det \bar{\psi }\left( 1-{{\gamma }_{5}} \right)\psi  \right].
\end{split}
\end{equation}
The effective potential with \textrm{AMM} can be taken as
\begin{equation}\label{eq:016}
	\begin{split}
	 {{\Omega }_{\text{AMM}}}=& {{G}_{s}}\sum\limits_{f=u,d,s}{\left\langle \overline{\psi }\psi  \right\rangle }_{f}^{2}+4K{{\left\langle \overline{\psi }\psi  \right\rangle }_{u}}{{\left\langle \overline{\psi }\psi  \right\rangle }_{d}}{{\left\langle \overline{\psi }\psi  \right\rangle }_{s}}
	-\frac{{N}_{c}}{2\pi } \sum\limits_{f=u,d,s}{\left| {{q}_{f}}B \right|} \sum\limits_{l=0}^{\infty } \sum\limits_{t=\pm 1} {\mathop{\int }_{-\infty }^{\infty }}\frac{d{{p}_{z}}}{2\pi }  \\
& \times \left\{ {{E}_{f,l,t}}+T\ln \left[ 1+\exp \left( \frac{-{{E}_{f,l,t}}-\mu }{T} \right) \right]+T\ln \left[ 1+\exp \left( \frac{-{{E}_{f,l,t}}+\mu }{T} \right) \right] \right\},
\end{split}
\end{equation}
where
\begin{equation}\label{eq:017}
{{E}_{_{f,l,t}}}=\sqrt{p_{z}^{2}+{{\left( {{\left( {{M}_{f}}^{2}+2|{{q}_{f}}B|l \right)}^{1/2}}-t{{\kappa }_{f}}{{q}_{f}}eB \right)}^{2}}}
\end{equation}
is the energy spectrum under different Landau energy levels, and $t=\pm 1$  corresponds to the two kinds of the spin direction of the $\bar{q} q $ pair. One can obtain three coupling gap equations for each order parameter as
\begin{equation}\label{eq:018}
	\frac{\partial {{\Omega }_{\textrm{AMM}}}}{\partial {{M}_{f}}}=0,
\end{equation}
where $f = u, d, s$ are for the three different flavors. Thus we can obtain three dynamical quark masses of $u, d$, and $s$ as
\begin{equation}\label{eq:019}
	\begin{split}
	& {{M}_{u}}={{m}_{u}}-4{{G}_{s}}{{\left\langle \bar{\psi }\psi  \right\rangle }_{u}}+2K{{\left\langle \overline{\psi }\psi  \right\rangle }_{d}}{{\left\langle \overline{\psi }\psi  \right\rangle }_{s}}, \\
	& {{M}_{d}}={{m}_{d}}-4{{G}_{s}}{{\left\langle \bar{\psi }\psi  \right\rangle }_{d}}+2K{{\left\langle \overline{\psi }\psi  \right\rangle }_{u}}{{\left\langle \overline{\psi }\psi  \right\rangle }_{s}}, \\
	& {{M}_{s}}={{m}_{s}}-4{{G}_{s}}{{\left\langle \bar{\psi }\psi  \right\rangle }_{s}}+2K{{\left\langle \overline{\psi }\psi  \right\rangle }_{u}}{{\left\langle \overline{\psi }\psi  \right\rangle }_{d}},
\end{split}
\end{equation}

\noindent where
\begin{equation}\label{eq:020}
{{\left\langle \bar{\psi }\psi  \right\rangle }_{f}}=\frac{{{N}_{c}}{{G}_{s}}}{2\pi }\underset{l=0}{\overset{\infty }{\mathop \sum }}\,{{\alpha }_{l}}|{{q}_{f}}B|\mathop{\int }_{-\infty }^{+\infty }\frac{d{{p}_{z}}}{2\pi }\frac{{{M}_{f}}}{{{\varepsilon }_{f,l,t}}}\left( 1-\frac{s{{\kappa }_{f}}{{q}_{f}}B}{{{{\hat{M}}}_{f,l,t}}} \right)\left\{ 1-\frac{1}{{{e}^{\frac{{{\varepsilon }_{f,l,t}}+\mu }{T}}}+1}-\frac{1}{{{e}^{\frac{{{\varepsilon }_{f,l,t}}-\mu }{T}}}+1} \right\}
\end{equation}

\noindent corresponds to the chiral condensation of different quark flavors.

\section{RESULTS AND DISCUSSIONS}\label{sec:03 contents}
To calibrate sets of parameters to applicable observables, parameters are chosen as $\Lambda =631.4~\mathrm{MeV}$, ${{m}_{u}}={{m}_{d}}=5.6~\mathrm{MeV}$, ${{m}_{s}}=135.7~\mathrm{MeV}$, ${{\Lambda }^{2}}{{G}_{s}}=1.835$ and $K{{\Lambda }^{5}}=9.29$~\cite{Hatsuda:1994pi,Kohyama:2016fif} . The empirical values are given as ${{f}_{\pi }}=93~\mathrm{MeV}$, ${{m}_{\pi }}=138~\mathrm{MeV}$, ${{m}_{K}}=495.7~\mathrm{MeV}$, and ${{m}_{\eta '}} = 957.5~\mathrm{MeV}$.

The tensor channel coupling constant $G_{t}$ restricted by the magnetic fields ought to be zero in the case of the vanished magnetic field, and equals the value of $G_{s}$ when $eB\to \infty $. In the following study, the value of $G_{t}$ is taken as $G_{t}=G_{s}/2$.

\begin{figure}[H]
	\centering
	\includegraphics[width=0.50\textwidth]{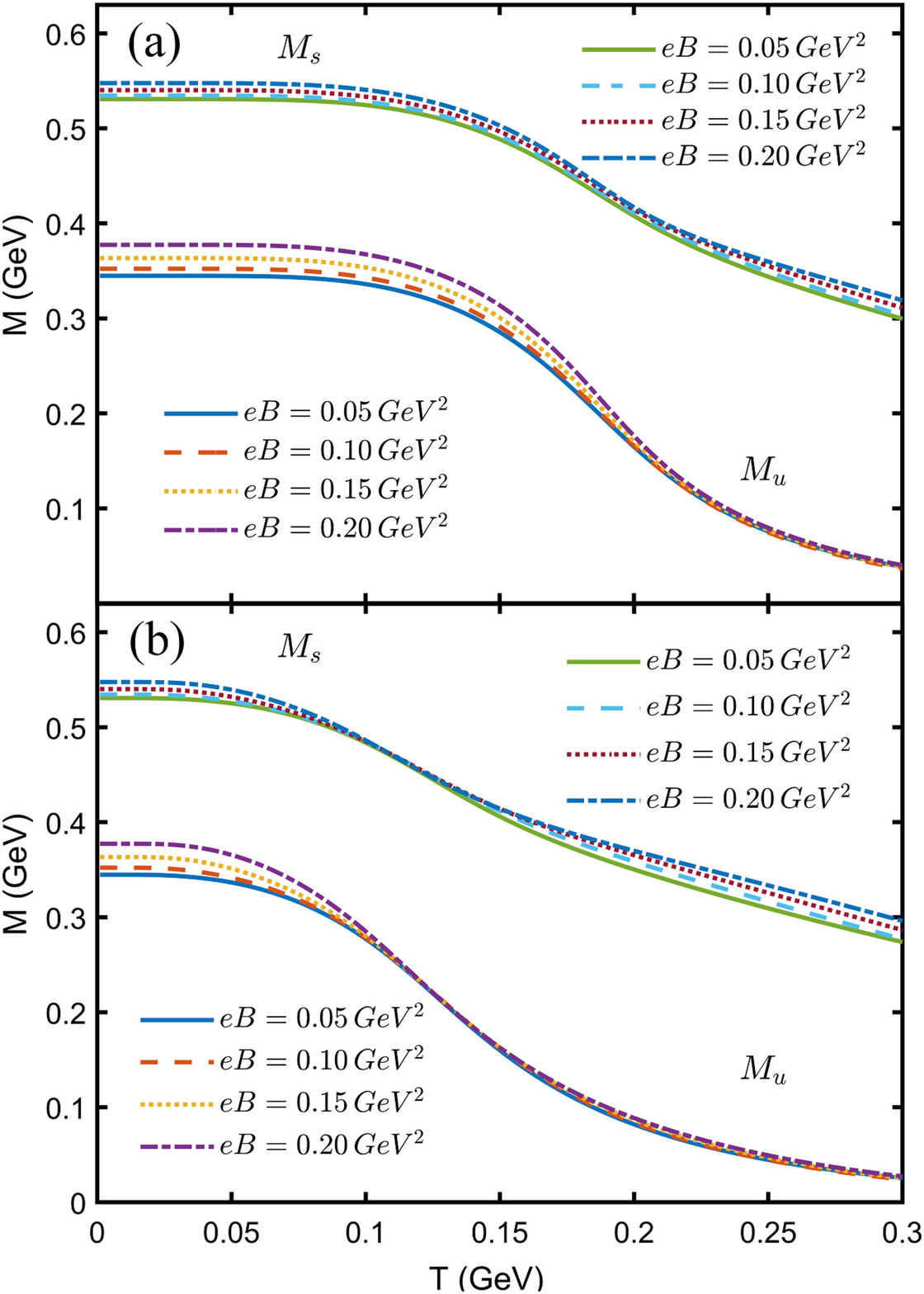}
	\caption{\label{fig1} The dependence of dynamical quark mass (M) on temperature (T)
		for four different magnetic fields ( $eB$ = 0.05, 0.10, 0.15 and 0.20 $\mathrm{GeV}^{2}$  ) with no considering \textrm{TSP} and \textrm{AMM}.
		(a) is for $\mu =0.0~\mathrm{GeV}$; and (b) is for $\mu =0.25~\mathrm{GeV}$.}
\end{figure}

In order to investigate the effect of \textrm{AMM} on the phase transition, we make comparisons between the two \textrm{AMM} sets. The compatible results obtained in \cite{Mekhfi:2005pn} we define it as AMM1 set as ${{\kappa }_{u}}={{\kappa }_{d}}=0.38,\ {{\kappa }_{s}}=0.25$, while the defined AMM2 set chosen as ${{\kappa }_{u}}=\text{0}\text{.123,} ~{{\kappa }_{d}}=0.555,\ {{\kappa }_{s}}=0.329$ fixed by \cite{Dothan:1981ex}.

Due to the \textrm{NJL} model being nonrenormalizable, the divergent vacuum terms merged in gap equation are regularized  by using the magnetic-field-independent regularization scheme \cite{Menezes:2008qt,Aguirre:2020tiy}, which gets rid of the nonphysical part by separating the vacuum term from the integrals. The scheme dealing with the sums of all Landau levels within the integrals by means of the Hurwitz zeta function is presented.

\begin{figure}[H]
	\centering
	\includegraphics[width=0.45\textwidth]{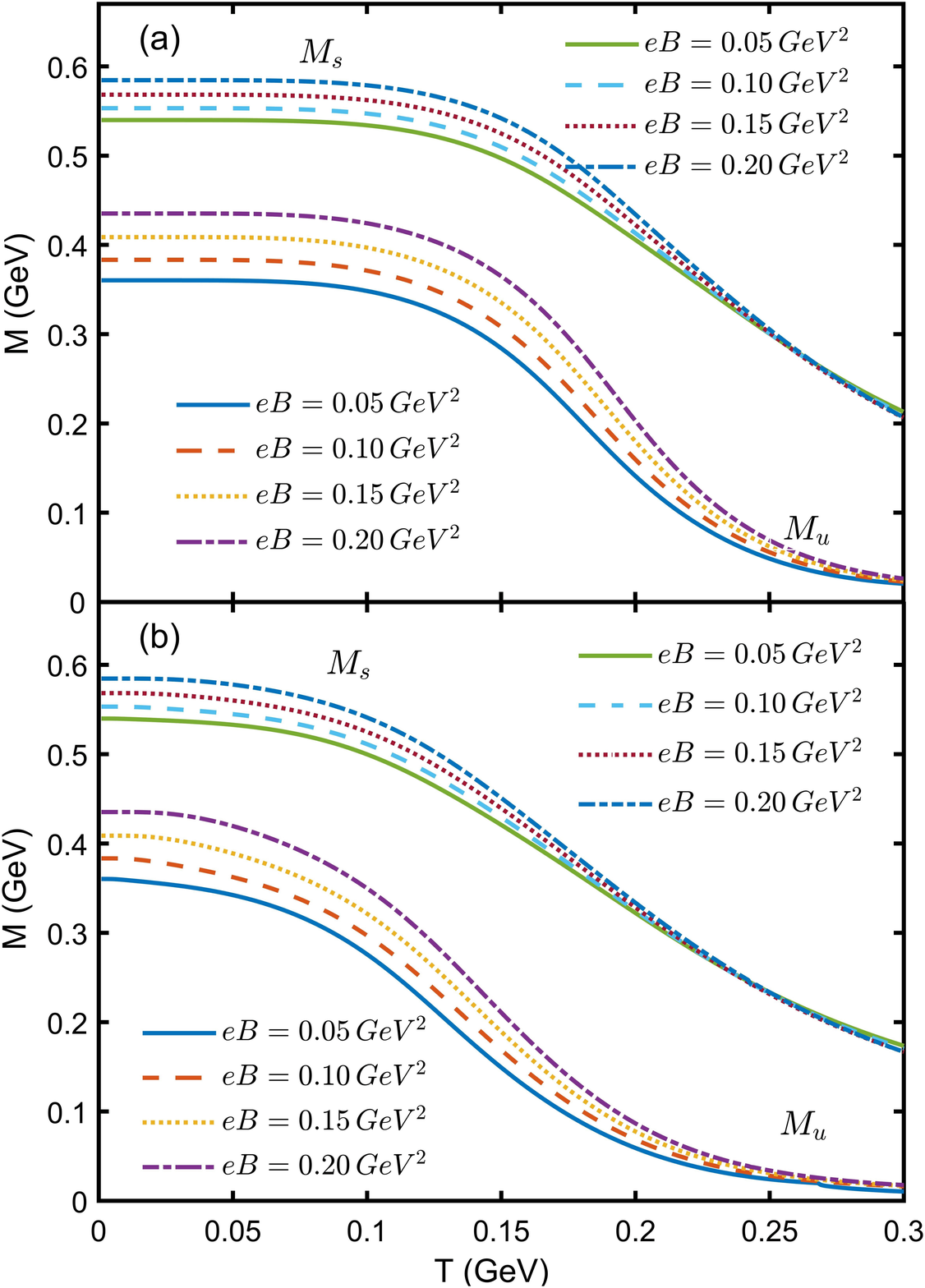}
	\caption{\label{fig2} The dependence of dynamical quark mass (M) on temperature (T)
		for four different magnetic fields ($eB$ = 0.05, 0.10, 0.15 and 0.20 $\mathrm{GeV}^{2}$)
		by considering \textrm{TSP}.  (a) is for $\mu =0.0~\mathrm{GeV}$  and (b) is for $\mu =0.25~\mathrm{GeV}$.}
\end{figure}
The dynamical mass or the quark condensate plays as an order parameter for the chiral- phase transition.  Chiral restoration
happens at high temperatures and/or high chemical potentials. In Fig. 1(a)-(b), the dynamical quark masses $M$ of $u$, $d$, and $s$ quarks without considering \textrm{AMM} and \textrm{TSP} are manifested as decreasing smooth functions of temperatures at $\mu =0~\mathrm{GeV}$ and $\mu =0.25~\mathrm{GeV}$, which indicates a chiral crossover. The dynamical mass $M$ is apparently enhanced by increasing the magnetic field. The magnetic field is shown at $eB$ = 0.05, 0.1, 0.15, and 0.2  $\mathrm{GeV}^{2}$ with $\mu =0~\mathrm{GeV}$ and $\mu =0.25~\mathrm{GeV}$, respectively. Since we have considered nonvanishing current quark mass, the chiral symmetry is never restored fully. Since the dynamical mass is proportional to chiral condensate, it can be seen from Fig. 1 that the larger the magnetic field is, the larger the corresponding chiral condensation is. This phenomenon is manifested as magnetic catalysis \cite{Kharzeev:2007jp,Gusynin:1999pq,Gusynin:1995nb,Gusynin:1994re}, which accounts for the magnetic field has a strong tendency to enhance (or catalyze) spin-zero $\bar{q} q$ condensates.

By considering \textrm{TSP} of quarks, we investigate the temperature dependence of constituent quark mass in Fig.2 (a)-(b) for $eB$ = 0.05, 0.10, 0.15 and 0.20 ~$\mathrm{GeV}^{2}$, respectively. The dynamical mass $M$ by considering \textrm{TSP} of quarks is manifested as a decreasing smooth function of temperatures for different magnetic fields and chemical potentials, which correspond to a chiral crossover. \textrm{TSP} is introduced by the anisotropic Fierz identity in \textrm{NJL} model, and moreover its essence is generated by symmetry broken caused by magnetic field $B$. The two tensor condensates $\left\langle \overline{\psi }{{\Sigma }_{3}}{{\lambda }_{3}}\psi  \right\rangle$  and $\left\langle \overline{\psi }{{\Sigma }_{3}}{{\lambda }_{8}}\psi  \right\rangle$  corresponding to \textrm{TSP} will provide a nonzero magnetic moment for the quasiparticle when the quark obtains the dynamical mass. The magnetic moment generated by the spin polarization under the action of the magnetic field will increase the dynamical mass of the quasiparticle, which leads to the MC effect. This MC characteristic of $u$ and $d$ quarks is more obvious in the high-temperature region.

\begin{figure}[H]
	\centering
	\includegraphics[width=0.85\textwidth]{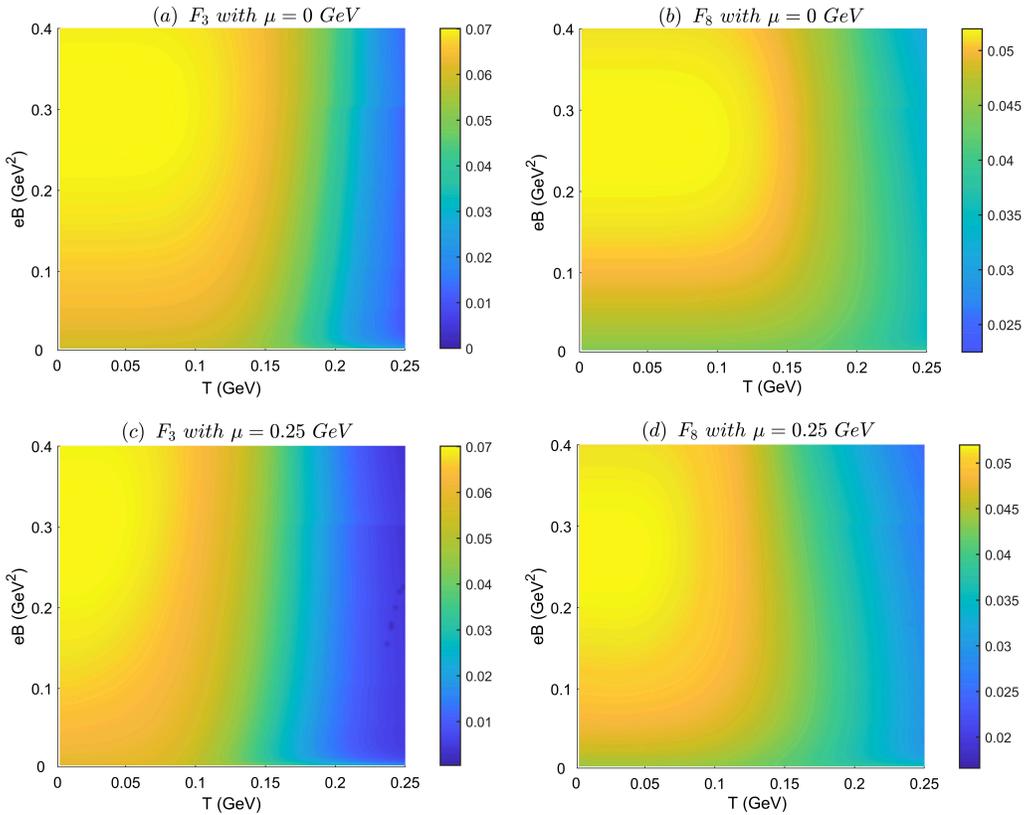}
	\caption{\label{fig3} Fig. 3(a),(b) shows the contour plots of the ${{F}_{3}}$ and ${{F}_{8}}$ distributions
		with zero chemical potential in the $T - eB$ plane, and Fig. 3(c),(d)
		shows similar plots of the ${{F}_{3}}$ and ${{F}_{8}}$ distributions but with nonzero
		chemical potential $\mu =0.25~\mathrm{GeV}$.}
\end{figure}

 In the $T- eB$  plane of Fig. 3, the corresponding temperature range is ~$0~\mathrm{GeV} \le T\le 0.3~\mathrm{GeV}$, and the magnetic field range is $0 ~\mathrm{GeV}^{2}\le eB\le 0.5\mathrm{GeV}^{2}$. Figure 3(a)and 3(b) displays the contour plots of the $F_{3}$ and $F_{8}$ distributions with a zero chemical potential in the $T- eB$ plane, and Figure 3(c)and 3(d) shows similar plots of the $F_{3}$ and $F_{8}$ distributions but with non-zero chemical potential $\mu = 0.25~\mathrm{GeV}$. The (2 + 1)-flavor spin polarization is different from that of two flavor spin polarization because of an additional term $F_{8} = -2G_{t}\left\langle \bar{\psi} \Sigma^{3}\lambda_{8}\psi  \right\rangle$  associated with the $\lambda_{8}$ flavor generator.

Figure 3 shows that both $F_{3}$ and $F_{8}$ become stronger at low temperatures, especially with the increase of the magnetic field. $F_{3}$ is almost zero at high temperature, and $F_{8}$ is very small but not zero at high temperature. The polarizations become weak at high temperatures(chiral symmetry phase restored area). It thus can be concluded that it is  more difficult to be polarized in the hot QGP background.

 \begin{figure}[H]
	\centering
	\includegraphics[width=0.75\textwidth]{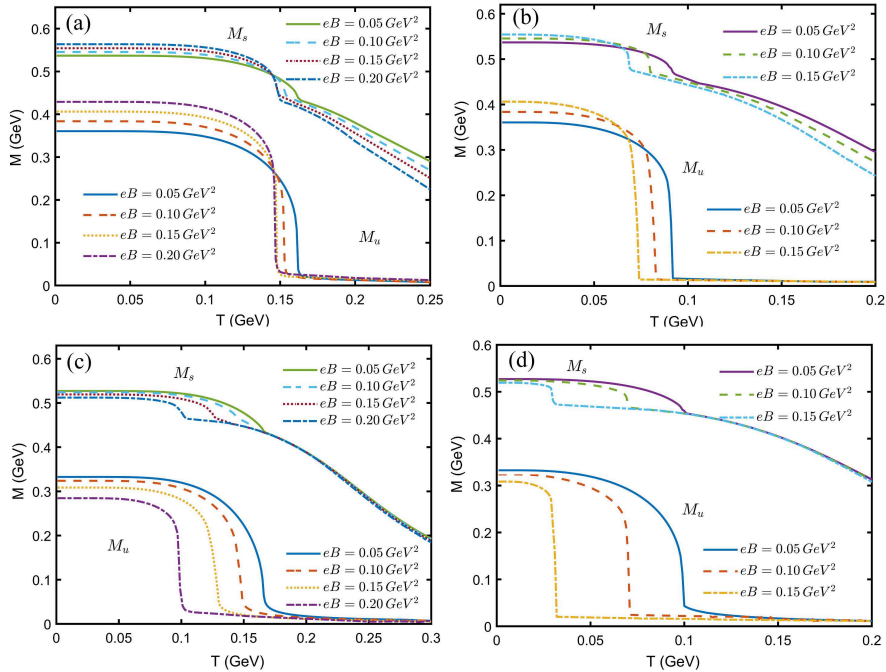}
	\caption{\label{fig4} The dynamical quark mass (M) as a function of temperature ($T$) for four different magnetic fields ($eB$ = 0.05, 0.10, 0.15 and 0.20 $\text{Ge}{{\text{V}}^{2}}$) by considering the different sets of \textrm{AMM}. Fig.4 (a, b) are for $\mu =0$ and $\mu =0.25 ~\mathrm{GeV}$ respectively with AMM1 set as $\kappa _{u}=\kappa_{d} = 0.38$, $\kappa_{s} = 0.25$. Fig.4(c, d) is same as Fig.4 (a, b) but for AMM2 set as $\kappa_{u}=0.123, \kappa_{d} = 0.555$, $\kappa_{s}=0.329$.}
\end{figure}

Figure 4. displays the dependence of dynamical quark mass ($M$) on temperature ($T$) for four different magnetic fields ($eB$ = 0.05, 0.10, 0.15 and 0.20 $\mathrm{GeV}^{2}$) by considering the two AMM's sets. Fig. 4(a, b) are for $\mu =0~\mathrm{GeV}$ and  $\mu =0.25 ~\mathrm{GeV}$ with AMM1 set as $\kappa _{u}=\kappa_{d} = 0.38$ and ${{\kappa }_{s}}=0.25$. Fig. 4(c, d) is same as Fig. 4(a, b) but with AMM2 set as $\kappa_{u}=0.123, \kappa_{d} = 0.555$ and $\ {{\kappa }_{s}} = 0.329$. Contrary to the behavior of the zero AMM in Fig. 1, the mass-decreasing behavior of $u$ and $d$ quarks in the chiral restoration is not a smooth slope but a sudden drop, which indicates the existence of a first-order transition.  However, the smooth slope of the dynamical mass for the crossover can be still present in the weak field $eB$ = 0.05 $\mathrm{GeV}^{2}$ for the non-zero AMM. The mass-decreasing behavior of $s$ quark in the chiral restoration is still a smooth slope, which suggests a chiral crossover for $s$ quark. From Fig. 4, it is found that the dynamical quark mass of $u$ and $d$ quarks have the characteristics of inverse magnetic catalysis in the chiral restoration phase ($T\ge T_{\textrm{C}}$) by using the \textrm{AMM} sets.

The generation of dynamical quark mass from the dimensional reduction from $3 + 1$ D to $1 + 1$ D is predominated by LLL at the low temperature region. That effect can be reflected in AMM1 in Fig. 4 obviously. More particles will be excited from LLL to a higher Landau level (LL) with the increasing temperature. The contribution of particles on higher LL to the dynamical mass by considering the effect of the AMM will decrease with the increasing of magnetic field, leading to the inverse magnetic catalytic characteristics. If the role of the AMM item is enough to alter the nature of the medium in the $T$ = 0 like the case of AMM2, the IMC effect characteristics of the AMM2 will be more significant.

\begin{figure}[H]
	\centering
	\includegraphics[width=0.85\textwidth]{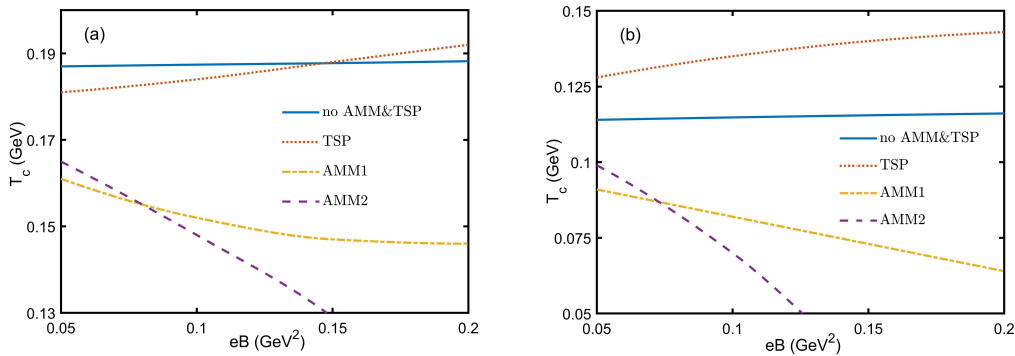}
	\caption{\label{fig5} The critical temperature of $u$ and $d$ quarks as a function of the magnetic field at $\mu$ = 0 (a) and $=0.25~\mathrm{GeV}$ (b).}
\end{figure}

In Fig. 5, the critical temperature is shown as a function of the magnetic field with the chemical potentials $\mu$ = 0 and $0.25~\mathrm{GeV}$, respectively. It is thus found that the critical temperature decreases with the magnetic field for the AMM1 and AMM2 sets, which indicates an inverse magnetic catalysis that qualitatively agrees with lattice result in~\cite{Bali:2012jv}. On the contrary, with the TSP, $T_{\textrm{C}}$ enhances as a function of the magnetic field, which is the extension of the magnetic catalysis effect from vacuum to finite temperature.
\begin{figure}[H]
	\centering
	\includegraphics[width=0.95\textwidth]{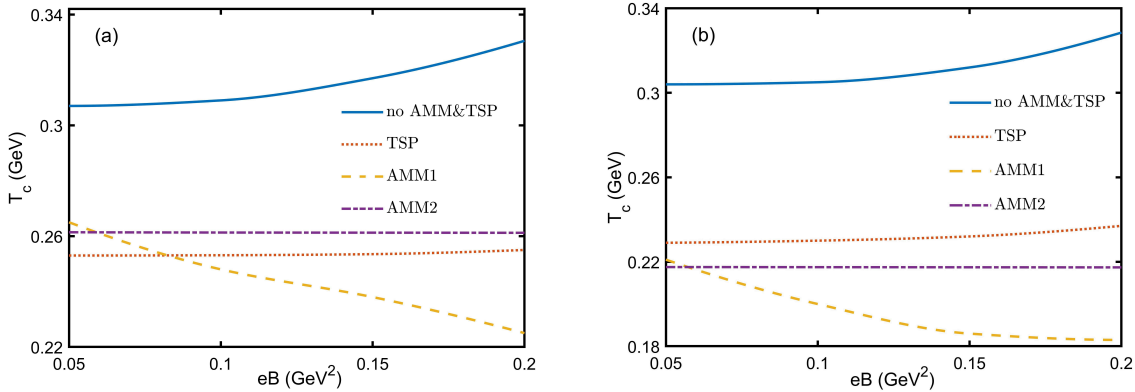}
	\caption{\label{fig6} The same as Fig. 5, but for the $s$-quark.}
\end{figure}

The critical temperature of the chiral phase transition of $s$ quark as a function of $eB$ is manifested in Fig. 6. Compared with light quarks of $u$ and $d$, the phase transition temperature $T_{\textrm{C}}$ of $s$ quark with \textrm{TSP} increases significantly with the increase of magnetic field, which corresponds to the characteristics of magnetic catalysis. The introduction of AMM sets corresponds to inverse magnetic catalytic characteristics.

Figure 7 displays the dependencies of the entropy density of $u$, $d$ and $s$ quarks on the temperature at zero chemical potential. It can be noted that the introduction of the \textrm{AMM} makes the crossover phase transition sharp. It is worth noting that the \textrm{AMM} in Fig.7 corresponds to three different settings, which are AMM0, AMM1 and AMM2, respectively. AMM0 means that the \textrm{AMM} is not considered, that is, all $\kappa $ values in Eq. (17) are set to zero. AMM1 and AMM2 sets have been mentioned above. When $eB = 0.05~\mathrm{GeV}^{2}$, the magnetic field is not big enough to excite the effect on entropy.  When $eB = 0.2~\mathrm{GeV}^{2}$, some of the effects of the magnetic field on entropy for different AMM sets and TSP can be excited. It is found that the entropy shows a sharp change near the phase transition temperature after adding \textrm{AMM} sets, and this sharp change is more obvious with the magnetic field increases and chemical potential, showing a first-order phase characteristic. The change of entropy with the temperature near the phase transition temperature is relatively smooth after adding \textrm{TSP}, and it behaves like the crossover transition.

\begin{figure}[H]
	\centering
	\includegraphics[width=0.5\textwidth]{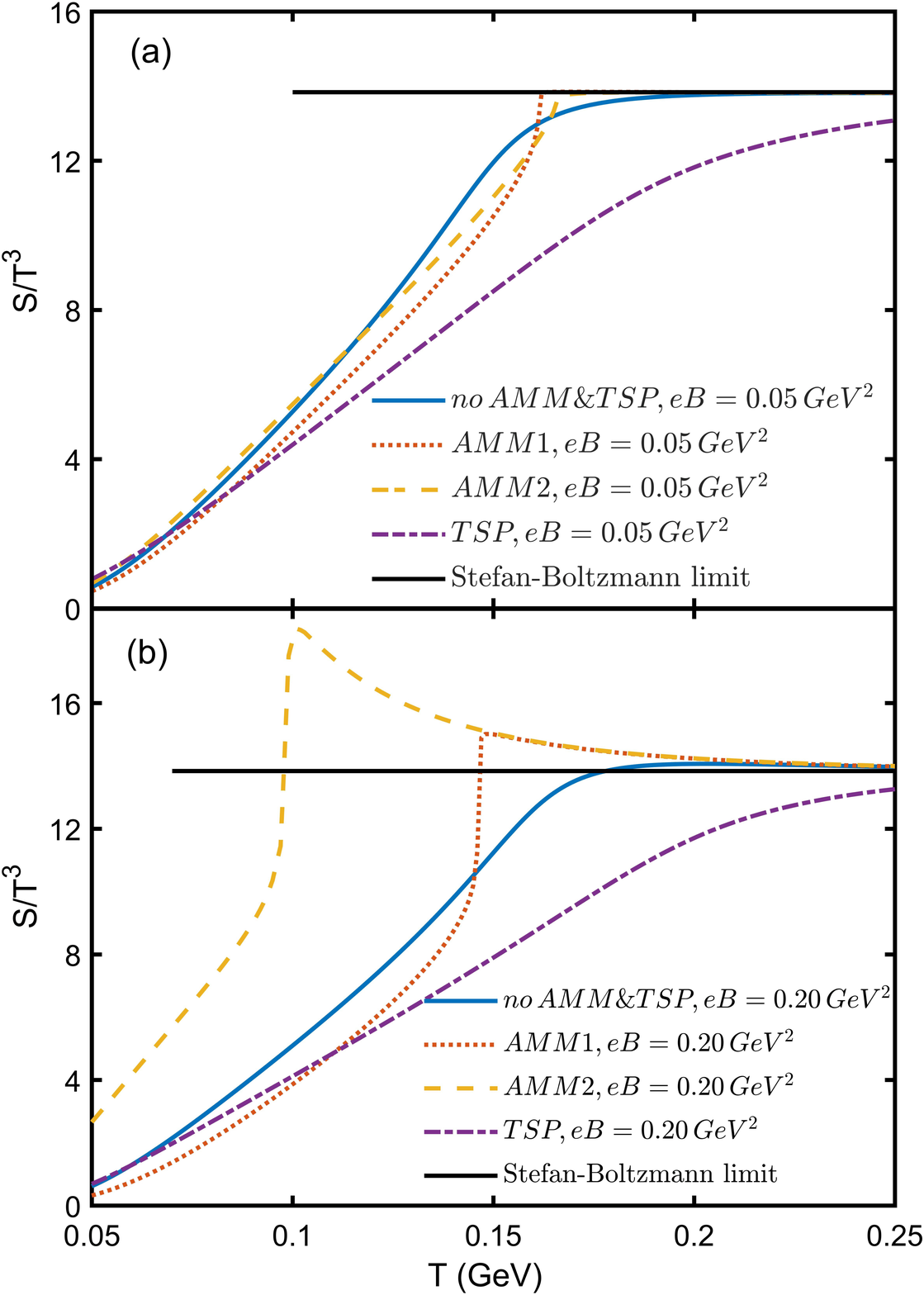}
	\caption{\label{fig7} The dependence of ${S}/{{{T}^{3}}}$ on temperature $T$ at $\mu = 0~ \mathrm{GeV}$ with different magnetic field.
		Fig. 7(a) is for $eB = 0.05~\mathrm{GeV}^{2}$ and Fig. 7(b) is for $eB = 0.2~\mathrm{GeV}^{2}$.}
\end{figure}

The dependence of square of sound-velocity $c_{\textrm{s}}^{2}$ on temperature $T$ is manifested in Fig. 8. Figure 8(a) and Fig. 8(b) are for zero chemical potential $\mu = 0 ~\mathrm{GeV}$ and $\mu = 0.25~\mathrm{GeV}$, respectively. In the region of crossover, the change of $c_{\textrm{s}}^{2}$  with temperature should be very smooth, while in the finite chemical potential region, the effect of \textrm{AMM} makes it incline to first-order transition. The bump rises rapidly because the dynamical quark mass has a discontinuous drop at this narrow region of temperature as shown in Fig. 4. Both sides of the narrow peak correspond to the hadron phase and quark-gluon plasma, and its highest point corresponds to the phase boundary. That leads our numerical result nearby the critical point of first-order transition to display nonphysical behavior even exceeding the Stefan-Boltzmann limit. Similar result after the consideration of \textrm{AMM} of two flavor has been reported in \cite{Wen:2021mgm}.

Compared with $u$ and $d$ quarks, the square of sound-velocity of $s$ quark with temperature is relatively smooth inflection after adding \textrm{TSP} and \textrm{AMM} sets.
It is proposed that $s$ quarks have always maintained obvious crossover characteristics. In the high-temperature region, the square of sound-velocity $c_{\textrm{s}}^{2}$ increases with temperature and obtains the saturation value $c_{\textrm{s}}^{2}=1/3$ to satisfy the relativistic requirement. This suggests that the equation of state in the chiral restoration phase at high temperatures is close to the Stefan-Boltzmann limit $\varepsilon = 3P$.
\begin{figure}[H]
	\centering
	\includegraphics[width=0.85\textwidth]{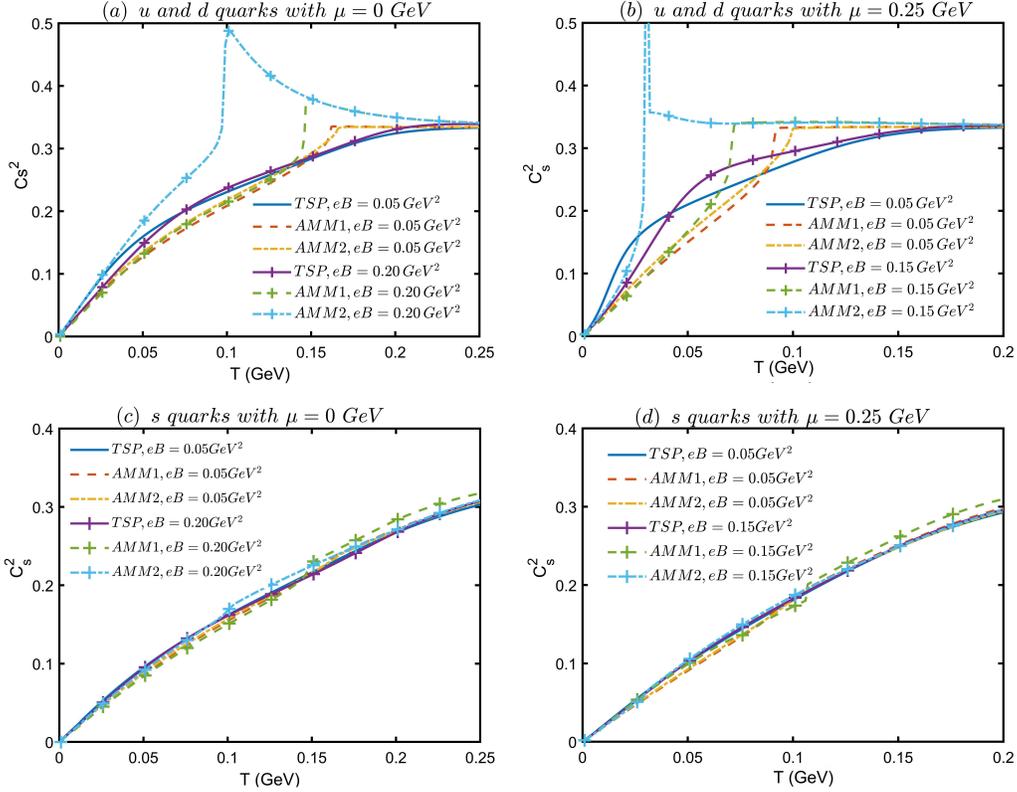}
	\caption{\label{fig8} The sound-velocity square $C_{\textrm{s}}^{2}$ of $u$ and $d$ with $s$ quarks as a function of the temperature $T$ with different chemical potential.
		(a), (b) is for $u$ and $d$ quarks with zero chemical potential $\mu$ = 0, and $\mu = 0.25~\mathrm{GeV}$, and (c),(d) is  for $s$ quarks}
\end{figure}

\section{SUMMARY AND CONCLUSIONS}\label{sec:04 contents}

In this work, we thoroughly study the effect from \textrm{TSP} and  \textrm{AMM} on the vacuum, phase transition and thermal magnetized \textrm{QCD} in the (2 + 1)-flavor Nambu-Jona-Lasinio (NJL) model with nonzero current quark masses at finite temperature and chemical potential. A unified physical mechanism to illustrate the novel consequences from recent lattice \textrm{QCD} as  magnetic catalysis and inverse magnetic catalysis effect proposed in the paper.

In the TSP case, since the dynamical quark mass is increased by the spin condensate, which is generated by an extra tensor channel independently as well as enhanced by the magnetic field, the pseudocritical temperature is increased by a rising magnetic field. This is why the magnetic catalysis feature appears in the case of \textrm{TSP}. While in the \textrm{AMM} case, the \textrm{AMM} term $\frac{1}{2}q_{f}\kappa\delta^{\mu\nu}F_{\mu\nu}$  does not directly produce a new condensate to impact the dynamical mass. Instead, it changes the energy spectrum of all Landau levels. As the result, it has been found that the \textrm{AMM} term will reduce the dynamical mass once the temperature is high enough to excite particles to jump to higher Landau levels.

It is found that the square of sound-velocity shows a sudden rapid rise in inflection near the phase transition
after adding \textrm{AMM} sets, and this rapid rise is more obvious with the magnetic field increases, showing an obviously first-order phase characteristic. On the other hand, after adding \textrm{TSP}, the change of
square of sound-velocity with temperature near the phase transition is relatively smooth inflection, showing an obviously crossover transition characteristic. The
result obtained by using the square of sound velocity is completely consistent with the result of entropy analysis.

The (2 + 1)-flavor spin polarization is different from that of two flavors because of an additional $F_{8} = -2G_{t}\left\langle \bar{\psi} \Sigma^{3}\lambda_{8}\psi  \right\rangle$ associated with the $\lambda_{8}$ flavor generator. The spin condensates affect the dynamical quark masses, chiral phase transition and quark dispersion relation. It is found that the polarizations become strong at low temperatures, and become weak at high temperatures. In other words, it is more difficult to be polarized in the hot \textrm{QGP} background, and easier to be polarized during the low-temperature region.

\section*{Acknowledgments}
This work was supported by the National Natural Science Foundation of China (Grants No. 11875178, No. 11475068, No. 11747115).

\section*{APPENDIX}
\setcounter{equation}{0}
\renewcommand\theequation{A.\arabic{equation}}
Following the regularization scheme to eliminate the divergent vacuum terms in the effective potential, the gap equation for $F_3$ and $F_8$ can be expressed as:

\begin{align}
	\begin{split}
			\frac{\partial \Omega}{\partial F_{3}} =
			&\textrm{FVac}_{u}+\textrm{FMag}_{u}+\textrm{FMed}_{u}+ \textrm{FVac}_{d}+\textrm{FMag}_{d}+\textrm{FMed}_{d} -F_3=0,
	\end{split}
\end{align}

\begin{align}
	\begin{split}
		\frac{\partial \Omega}{\partial F_{8}} =
		& \frac{1}{\sqrt{3} } \big( \textrm{FVac}_{u}+\textrm{FMag}_{u}+\textrm{FMed}_{u}-\textrm{FVac}_{d}-\textrm{FMag}_{d}-\textrm{FMed}_{d} \big)+\\
		&2\big (   \textrm{FVac}_{s}+\textrm{FMag}_{s}+\textrm{FMed}_{s}\big )-F_8=0,
	\end{split}
\end{align}
where the terms are:
\begin{align}
	\begin{split}
\textrm{FVac}_{f}=\frac{N_c}{2 \pi ^2} \ ( M_f + \textrm{MX}_f \ )\big( \Lambda \sqrt{\Lambda ^2+{M_f}^2} -\frac{{M_f}^2}{2}\ln_{}{\ ( \frac{\ (\Lambda +\sqrt{\Lambda ^2+{M_f}^2} \ )^2 }{{M_f}^2}  \ ) } \big),
	\end{split}
\end{align}
\begin{align}
	\begin{split}
		\textrm{FMag}_{f}=&\frac{N_c\ | q_feB\ |}{4 \pi ^2} \bigg\{ M_f \ln{x_{f,-1}}+
	\textrm{MX}_f \big( \ln{\ [ \Gamma \ ( x_{f,1} \ )  \ ] }+ \ln{\ [ \Gamma \ ( x_{f,-1} \ )  \ ]}+ \ln{2 \pi}    \\
		&-\ ( -x_{f,1}-x_{f,-1}+x_{f,-1}\ln{x_{f,-1}}+x_{f,1} \ln{x_{f,1}}  \ )     \big)  \bigg\},
	\end{split}
\end{align}
\begin{align}
	\begin{split}
		\textrm{FMed}_{f}=
		&\frac{N_c\ | q_feB\ |}{4 \pi ^2}\bigg\{ \int_{-\infty}^{\infty}dp_z \frac{\textrm{MX}_f+ M_f}{E_{f,l=0} }\bigg(  \frac{1}{1+ \exp\ ( \frac{E_{f,l=0}-\mu }{T} \ )} +\frac{1}{1+ \exp\ ( \frac{E_{f,l=0}+\mu }{T} \ )} \bigg) \\
		&+\sum_{l=1}^{\infty } \alpha _{l}\int_{-\infty}^{\infty} dp_z\frac{\textrm{MX}_f+\eta M_f}{E_{f,l,\eta} }\bigg(  \frac{1}{1+ \exp\ ( \frac{E_{f,l,\eta}-\mu }{T} \ )} +\frac{1}{1+ \exp\ ( \frac{E_{f,l,\eta}+\mu }{T} \ )} \bigg) \bigg\},
	\end{split}
\end{align}
where $x_{f,\eta}$ and $\textrm{MX}_f$ are given as
\begin{align}
	\begin{split}
&x_{f,\eta}=\frac{1}{2\ | q_feB\ |}\bigg({M_f}^{2}+{\textrm{MX}_f}^{2}+2\eta M_f \textrm{MX}_f \bigg),\\	
	\end{split}
\end{align}

\begin{align}
\begin{split}
&\textrm{MX}_u=\frac{F_3+F_8}{\sqrt{3} },\\
&\textrm{MX}_d=\frac{F_3-F_8}{\sqrt{3} },\\
&\textrm{MX}_s=\frac{2F_8}{\sqrt{3} }.
\end{split}
\end{align}
where the $\eta$ indicates the spin label $\pm$ 1.

\section*{References}

\bibliography{ref}

\end{document}